\def\year{2021}\relax
\documentclass[letterpaper]{article} 
\usepackage{aaai21}  
\usepackage{times}  
\usepackage{helvet} 
\usepackage{courier}  
\usepackage[hyphens]{url}  
\usepackage{graphicx} 
\usepackage{amsmath}
\usepackage{natbib}  
\usepackage{caption} 
\title{The Manufacture of Partisan Echo Chambers\\by Follow Train Abuse on Twitter}
\author{
    Christopher Torres-Lugo,
    Kai-Cheng Yang, 
    Filippo Menczer \\
}

\affiliations {
    Observatory on Social Media, Indiana University, Bloomington, USA\\
}

\begin{document}

\newcommand{\hashtag}[1]{{\textit{\##1}}}
\newcommand{\keyword}[1]{{\textit{``#1''}}}
\newcommand{\mention}[1]{{\textit{\@#1}}}
\newcommand{\dataset}[1]{{\texttt{#1}}}

\maketitle

\begin{abstract}
A growing body of evidence points to critical vulnerabilities of social media, such as the emergence of partisan echo chambers and the viral spread of misinformation.
We show that these vulnerabilities are amplified by abusive behaviors associated with  so-called ``follow trains'' on Twitter, in which long lists of like-minded accounts are mentioned for others to follow. 
We present the first systematic analysis of a large U.S. hyper-partisan train network. 
We observe an artificial inflation of influence: accounts heavily promoted by follow trains profit from a median six-fold increase in daily follower growth.  
This catalyzes the formation of highly clustered echo chambers, hierarchically organized around a dense core of active accounts. 
Train accounts also engage in other behaviors that violate platform policies: we find evidence of activity by inauthentic automated accounts and abnormal content deletion, as well as amplification of toxic content from low-credibility and conspiratorial sources. 
Some train accounts have been active for years, suggesting that platforms need to pay greater attention to this kind of abuse.
\end{abstract}

\section{Introduction}

In the past decade, online social media have become an important platform for participating in social movements regarding various public issues like economic inequality~\cite{gleason2013occupy,conover2013digital},
human rights~\cite{stewart2017drawing}, and especially political elections~\cite{flores2018mobilizing}. 
Features of social media like peer-to-peer communication, anonymity, high efficiency, and broad coverage greatly facilitate organization efforts~\cite{lotan2011arab,starbird2012will,tufekci2012social}.
Unfortunately, the same features invite problems like inauthentic behaviors~\cite{ferrara2016rise,pacheco2020uncovering}, echo chambers~\cite{sasahara2020social}, and the wide spread of toxic information~\cite{shao2018spread,grinberg2019fake,ahmed2020covid} that challenge the integrity of the information ecosystem. 
The January 6, 2021 riot at the U.S. Capitol provide clear evidence of the entanglement between online disinformation and real-world harm, making it all the more crucial to study how bad actors manipulate online discourse.\footnote{\url{apnews.com/article/donald-trump-conspiracy-theories-michael-pence-media-social-media-daba3f5dd16a431abc627a5cfc922b87}}

\begin{figure*}[t]
    \centering
    \includegraphics[width=\textwidth]{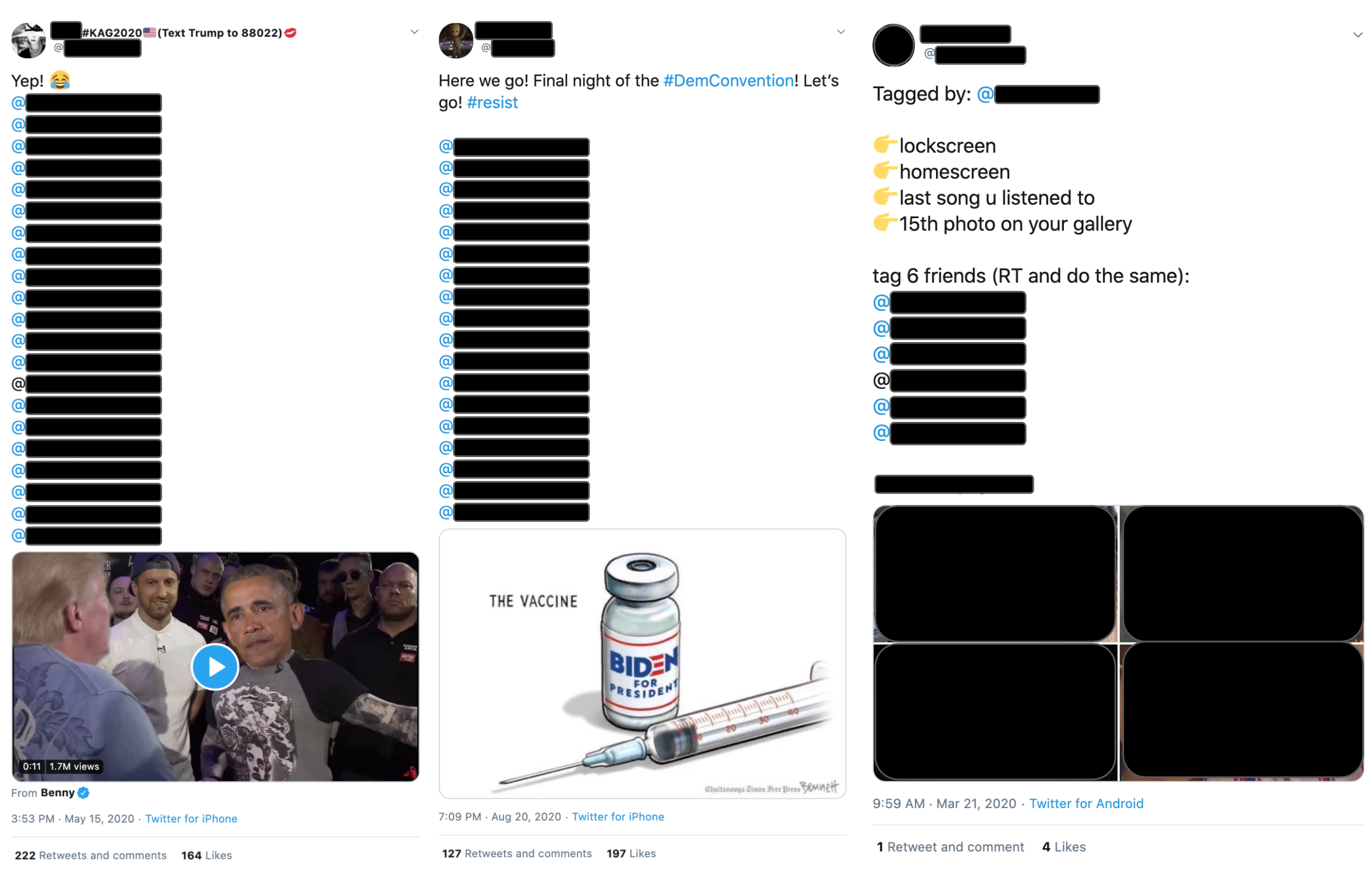}
    \caption{Screenshots of exemplar  tweets from (left) a pro-Trump train conductor, (center) an anti-Trump train conductor, and (right) the tagging game. Some account information is redacted to protect privacy.}
    \label{fig:screenshot}
\end{figure*}

``Follow trains'' are a way for social media users to suggest other accounts to their followers. 
Our own experience suggests that follow trains are widely abused for political manipulation on Twitter and other social media platforms. 
The characteristic behavior of partisan follow trains is the publishing of spam-like ``train tweets'' that typically contain a list of mentions (accounts), a media object, and possibly a few words.
The screenshots in Figure~\ref{fig:screenshot} show examples of two political follow train tweets. 
The organizers (``train conductors'') post the train tweets so that their followers can follow the accounts being mentioned (``train riders'').

The essential goal of conducting follow trains is for the riders to efficiently gain more followers, which is a direct violation of Twitter's platform manipulation and spam policy.\footnote{\url{help.twitter.com/en/rules-and-policies/platform-manipulation}}
Tweets from train accounts are used to constantly seek attention from government officials and politicians by means of mentioning, retweeting, and replying. Occasionally they get amplified by influential users and reach a much broader audience. For example, President Trump has retweeted train accounts in the past.\footnote{See \url{web.archive.org/web/20200410122951/https://twitter.com/realdonaldtrump/status/1248589007329595392} for an archived example (Trump's account is suspended as of this writing).}

Toward the goal of understanding partisan follow trains, we present, to the best of our knowledge, the first systematic analysis of such abuse.
We focus on the Twitter platform and specifically pro-Trump follow trains. We present an analytical characterization of train accounts from multiple perspectives, with an emphasis on their social networks and questionable behaviors. We build and share datasets of train accounts, collect their tweets, and contrast their behaviors against those of accounts in baseline datasets to provide meaningful contexts for our analysis. This paper contributes the following findings:
\begin{itemize}
    \item Pro-Trump follow trains are highly effective in inflating follower numbers for the train riders. 
    \item Train accounts are more active than baseline pro-Trump Twitter users in terms of establishing social ties. As a result, they form a hierarchical and dense community that is highly coordinated, persistent, homogeneous, and fully focused on amplifying conservative narratives. These are characteristics that we use to label the community as a partisan echo chamber~\cite{sasahara2020social}. 
    \item Train accounts are also more active than baseline pro-Trump users in terms of posting tweets. 
    \item In addition to generating and amplifying a large amount of spam-like tweets, some of the accounts abuse the platform through inauthentic personas and abnormal tweet deletions. 
    \item Finally, by analyzing their tweets, we find that train accounts  actively share a large volume of toxic information, such as low-credibility news and conspiracy theories.
\end{itemize}

\section{Background}

Social media influence stems from having high visibility on a platform. Despite early analysis showing that influence is not exclusively determined by the number of followers~\cite{cha2010measuring}, it is generally assumed that users need to have many followers to increase their visibility. Some unscrupulous actors therefore resort to follower growth hacking.
A well-studied growth hack is to purchase ``fake followers,'' which often consist of inauthentic or compromised accounts~\cite{cresci2015fame,aggarwal2015they}.
There are also reports of organized exchanges to turn unpaid customers and volunteers into fake followers~\cite{stringhini2013follow,liu2016pay}.  
Follow trains are a particularly effective follower growth hack: actions are coordinated with the ultimate goal of building a well-connected community with maximal influence.

In addition to violating Twitter's policies, partisan follow trains may also produce undesirable outcomes.
For example, polarized and segregated echo chambers are commonly observed on social media~\cite{Jamieson2008,garrett2009echo,Truthy_icwsm2011politics,lee2014social}, possibly leading to radicalization~\cite{wojcieszak2010don,Bright2016}. 
Behind the curtain is the interplay between social biases like the tendency to establish belief-consistent social ties~\cite{del2017modeling,HillsProliferation18}, cognitive biases such as information overload and confirmation bias~\cite{MenczerHills2020SciAm}, and social media mechanisms like friend recommendations and the ease of (un)following~\cite{sasahara2020social}. 
Blindly following riders recommended by like-minded conductors can easily accelerate the formation of polarized echo chambers.

Partisan follow trains also make the online community more vulnerable to inauthentic actors, who are blindly followed when recommended. 
Well-known types of inauthentic accounts include trolls~\cite{zannettou2019disinformation} and malicious social bots~\cite{ferrara2016rise}, which have been actively involved in  online discussions of elections across various countries~\cite{bessi2016social,deb2019perils,stella2018bots,ferrara2017disinformation,badawy2018analyzing,zannettou2019disinformation}. 
Follow trains provide an easy mechanism for an entity to control automated accounts programmed to follow accounts mentioned by a conductor. Even train conductors can be automated. 
Recently, more attention has been drawn toward a new type of inauthentic actors that act in a coordinated fashion to increase influence and evade detection~\cite{nizzoli2020coordinated,sharma2020identifying,pacheco2020uncovering}.
Follow trains facilitate the formation of coordinated inauthentic networks.

Another problem regarding follow trains is the spread of toxic information, such as conspiracy theories and misinformation.
Concerns about ``fake news'' on social media have been growing since the 2016 U.S. presidential election~\cite{Lazer-fake-news-2018,vosoughi2018spread,grinberg2019fake,bovet2019influence}.
During the 2020 COVID-19 pandemic, misinformation related to the outbreak, also known as the ``infodemic,'' has also spread virally~\cite{zarocostas2020fight,yang2020prevalence, yang2020covid}. 
Recent studies show that polarized echo chambers are associated with the diffusion of misinformation~\cite{del2016spreading} and inauthentic actors like malicious bots are responsible for spreading low-credibility information related to politics~\cite{shao2018spread}; this suggests that partisan follow trains may also exacerbate the misinformation problem --- a conjecture we explore in this paper.
Due to the potential real-world consequences of misinformation and conspiracy theories about topics such as health and elections, it's important to thoroughly investigate the role of partisan follow train in such abuse.

\section{Data Collection}

All data and code necessary to reproduce the results in this paper are shared in a public repository.

\subsection{Network Analysis}

First, we focus on the social (mention and follow) networks of train accounts.
We label a tweet as a \emph{follow train tweet} using a simple heuristic rule, namely, if it mentions nine or more other accounts and contains media. 
We selected the threshold on the number of mentioned accounts based on the observation that nine of them tend to occupy a majority of the tweet text, suggesting that the mentions are the main purpose of the tweet. 
Any account having at least one original follow train tweet is considered a \emph{train conductor}.
Accounts mentioned in follow train tweets are labeled as \emph{train riders}.
A conductor can also be a rider, but the conductor label takes precedence in our dataset.
Such an operationalization allows us to distinguish train conductors and riders at scale.

We utilize snowball sampling to crawl a mention network of partisan train conductors and riders.
Since the focus of the present paper is on pro-Trump follow trains, we start the crawl from a high-profile pro-Trump train conductor. 
We query the Twitter search API to retrieve this account's follow train tweets (retweets and replies are excluded). All the mentioned accounts (riders) are extracted. 
Since our goal is to focus on influential conductors, we exclude 14\% of train tweets that have been retweeted less than 40 times. 
The procedure is repeated recursively on the riders, until no new accounts emerge.
To further refine the dataset, we conduct an exhaustive manual annotation of the accounts collected.
We find 9\% of them are verified accounts, 3\% are suspended or deleted, and 1\% are non-political or inactive.
These accounts are removed from further analysis.
The remaining accounts are hyper-partisan. Among them we find a small group of anti-Trump accounts. 
Further inspection reveals that this is due to a pro-Trump account mentioning an anti-Trump account.
The anti-Trump portion of the network is also excluded from the present analysis. 

The data collection took place between February 16 and 26, 2020.
The resulting mention network, denoted as \dataset{train-net}, contains 8,308 nodes (182 conductors and 8,126 riders) and 20,773 edges.
Note that this categorization only reflects the behaviors of the accounts in the data collection period; as mentioned earlier, riders could act as conductors at a different time.
Figure~\ref{fig:wordcloud} shows common hashtags in the profile descriptions of the train accounts in the network, demonstrating a clear alignment with pro-Trump themes.

To gauge the structure of the mention network, we need a suitable baseline. We use a similar method to collect a group of accounts that share similar behaviors but in a non-political context.
We take advantage of an online game played by many Twitter users in the early days of the 2020 COVID-19 lockdown.
In the game, each tagged user is asked to post a screenshot of their phone and mention six friends to continue the game with the same instruction.
An exemplar tagging tweet can be seen in Figure~\ref{fig:screenshot}.
We start from a participant and search for more by crawling the mention network.
We identify game participants through phrases like \keyword{tagged by,} \keyword{lockscreen,} \keyword{homescreen,} and \keyword{last song u listened to}; other collected accounts are removed.
The data collection took place between March 21 and 27, 2020.
The resulting mention network has 5,567 nodes and 7,189 edges.
We denote this dataset as \dataset{tagging-net}.

\subsection{Behavioral Analysis}

The small number of conductor accounts in the \dataset{train-net} dataset makes it difficult to obtain a robust statistical analysis of their profiles and behaviors. 
To expand the number of conductors, we leverage the observation that accounts engaged in train behavior retweet follow train tweets by other conductors.
We selected all the conductors and a similarly-sized random sample (18\%) of the riders in \dataset{train-net}. The user IDs of these accounts were used to query a historical archive that offers the so-called Decahose, a 10\% random sample of the public tweet stream
~\cite{osome}. 
We scanned the resulting sample of  tweets, retweets, and quotes posted by the selected train accounts between January 1, 2018 and June 30, 2020 for additional train tweets meeting our heuristic rules. 
The accounts that published those train tweets were considered to be additional conductors, excluding verified accounts.
We finally retrieved a sample of tweets by the additional conductors as well, in the same period, again using the historical archive.

\begin{figure}[t]
    \centering
    \includegraphics[width=0.85\columnwidth]{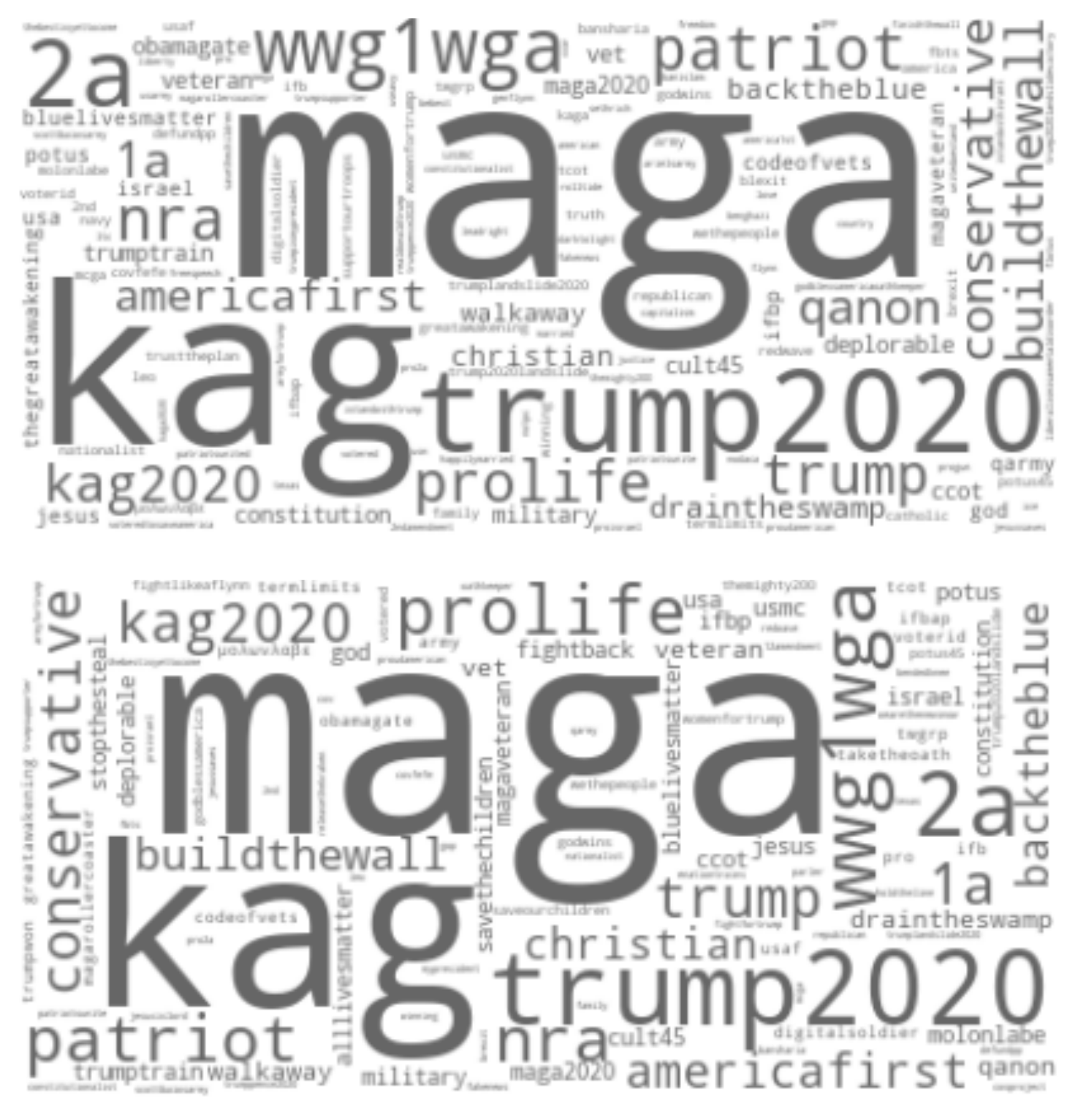}
    \caption{Frequent hashtags in pro-Trump account descriptions in (top) \dataset{train-net} and (bottom) \dataset{decahose}. 
    }
    \label{fig:wordcloud}
\end{figure}

While the train conductors and riders in our data are all pro-Trump, not all pro-Trump accounts are involved in follow trains.
We therefore need another suitable baseline to examine whether the characteristics of the train accounts are typical of pro-Trump users. 
We build a representative sample of pro-Trump Twitter accounts starting from a list of 
hashtags frequently found in pro-Trump account descriptions from \dataset{train-net} (see Figure~\ref{fig:wordcloud}). 
We scan the historical archive and find all the accounts that tweeted on February 20, 2020 and that had any of the selected hashtags in their descriptions.
Among the matched partisan accounts, we further sample users to approximately match the number of train accounts.
As for the pro-Trump train accounts, we collect historical tweets by the pro-Trump baseline accounts between January 1, 2018 and June 30, 2020.

We denote the historical dataset containing all the conductors (those from \dataset{train-net} plus the additional ones), the sampled riders, and the baseline as \dataset{decahose}. It contains 14,693,650 tweets by 4,581 pro-Trump accounts: 1,342 conductors, 1,352 riders, and 1,839 baseline accounts. Figure~\ref{fig:wordcloud} shows the most frequent hashtags in the descriptions of these accounts.
A clear partisan alignment  with \dataset{train-net} accounts can be observed.

\section{Anatomy of Follow Train Networks}

\subsection{Mention Networks}

\begin{figure}[t]
    \centering
    \includegraphics[width=\columnwidth]{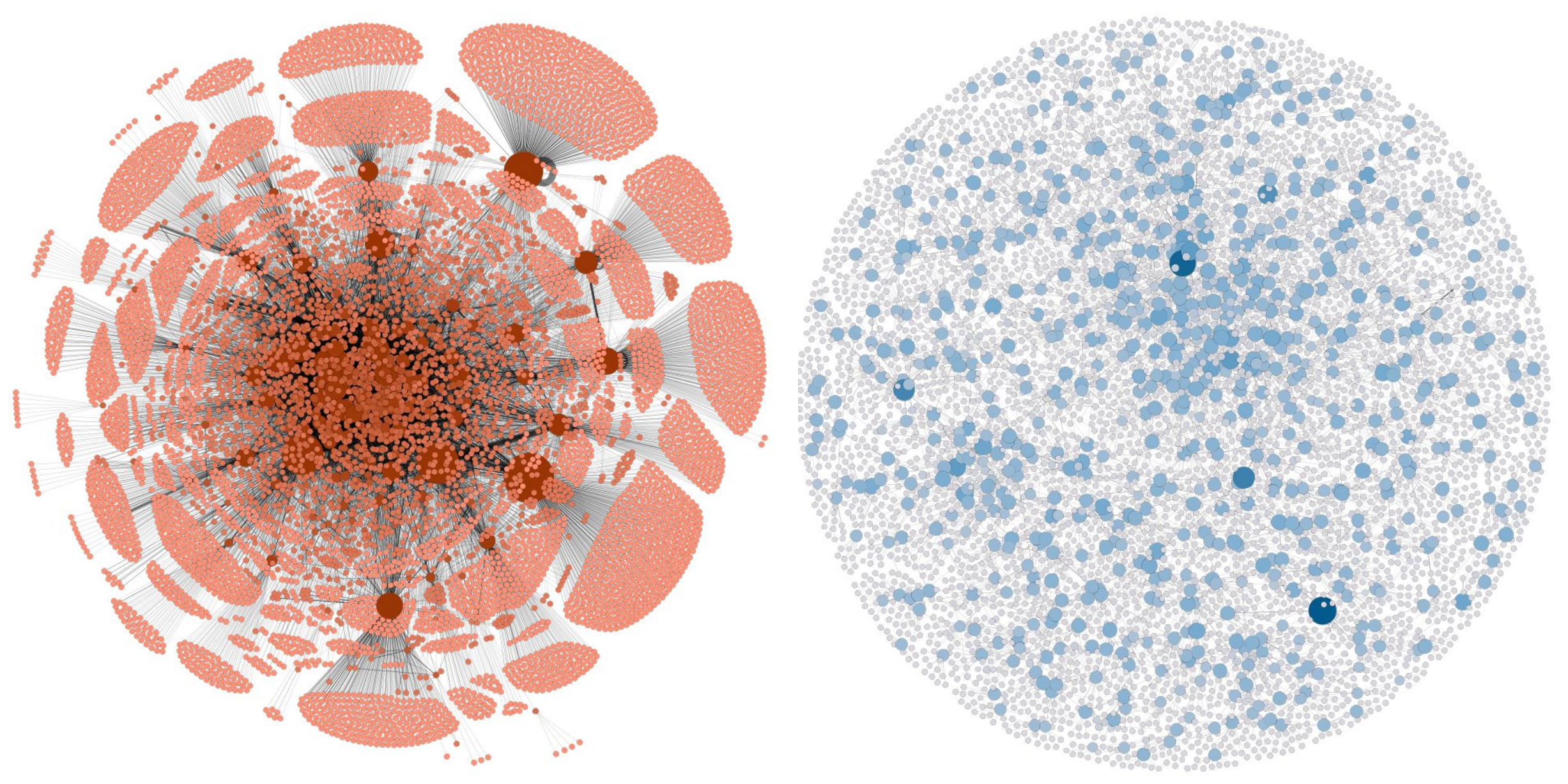}
    \caption{Visualizations of the mention networks among accounts in the (left) \dataset{train-net} dataset and (right) \dataset{tagging-net} baseline. We used the Fruchterman-Reingold layout algorithm in Gephi with identical parameters for both networks. Larger size and darker color of a node denote higher degree. In the train network, singleton nodes (riders mentioned by suspended conductor accounts) are not shown.
    }
    \label{fig:mentioning_network}
\end{figure}

Let us first analyze the \dataset{train-net} and \dataset{tagging-net} mention networks, in which each node represents an account and an edge from node A to B means that account A has mentioned B. Edge weights represent the numbers of mentions.
Figure~\ref{fig:mentioning_network} visualizes the networks, showing a much more hierarchical structure in the follow train network compared to the baseline. 

\begin{figure}[t]
    \centering
    \includegraphics[width=\columnwidth]{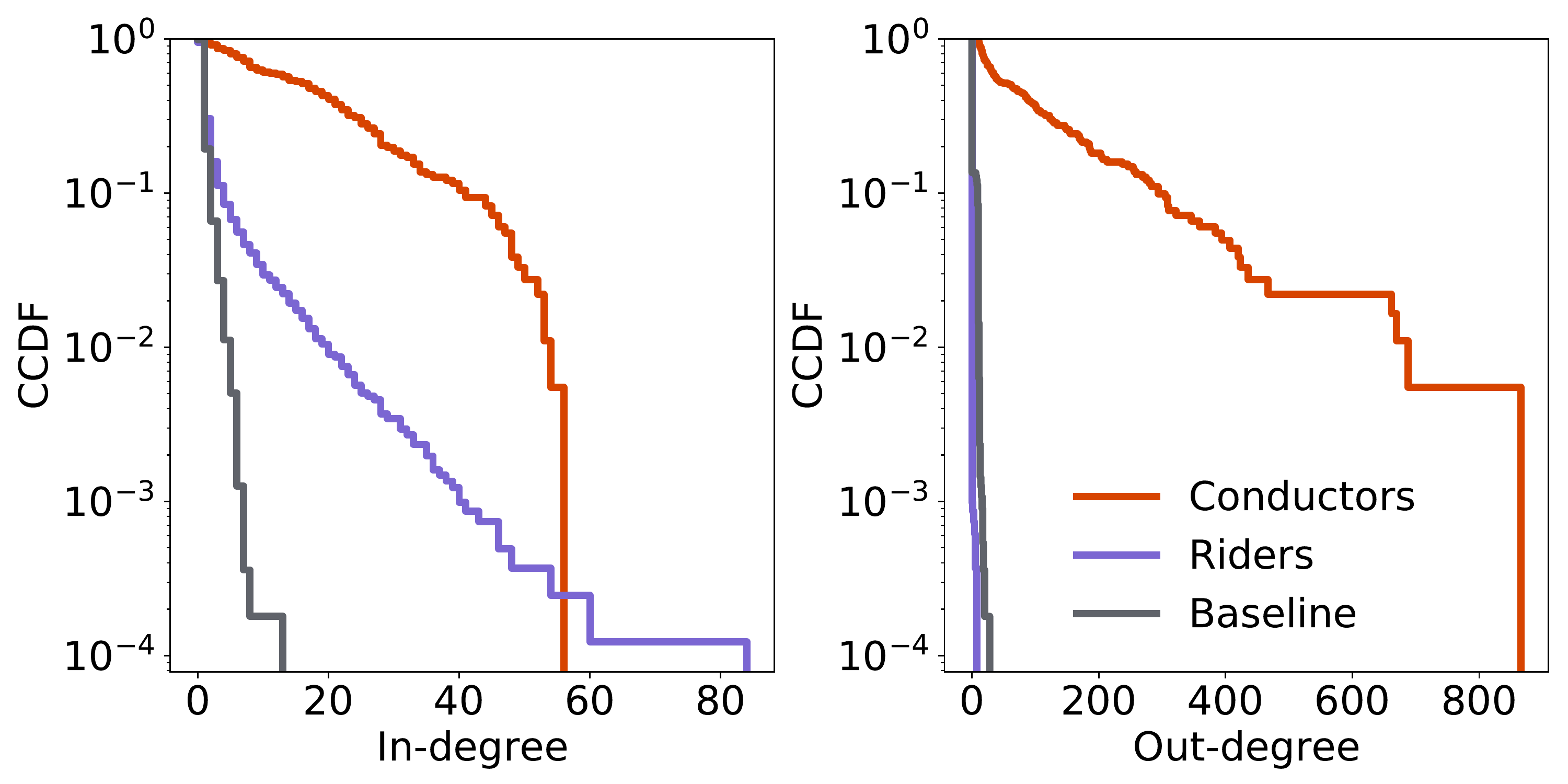}
    \caption{Complementary cumulative distributions of (left) in-degree and (right) out-degree for accounts in the  \dataset{train-net} mention network and \dataset{tagging-net} baseline.
    Kolmogorov–Smirnov (KS) tests show that all of the distributions are significantly different from each other ($p<0.01$).}
    \label{fig:degree_distribution}
\end{figure}
    
To characterize the mention network, the in-degree and out-degree distributions of the nodes are plotted in Figure~\ref{fig:degree_distribution}.
High in-degree indicates that an account was mentioned by many others and high out-degree means the account mentioned many others.
We find that conductors tend to have much higher out-degree than rider and baseline accounts, consistent with their role. As a result, riders are mentioned more than baseline accounts. 

\begin{figure}[t]
    \centering
    \includegraphics[width=\columnwidth]{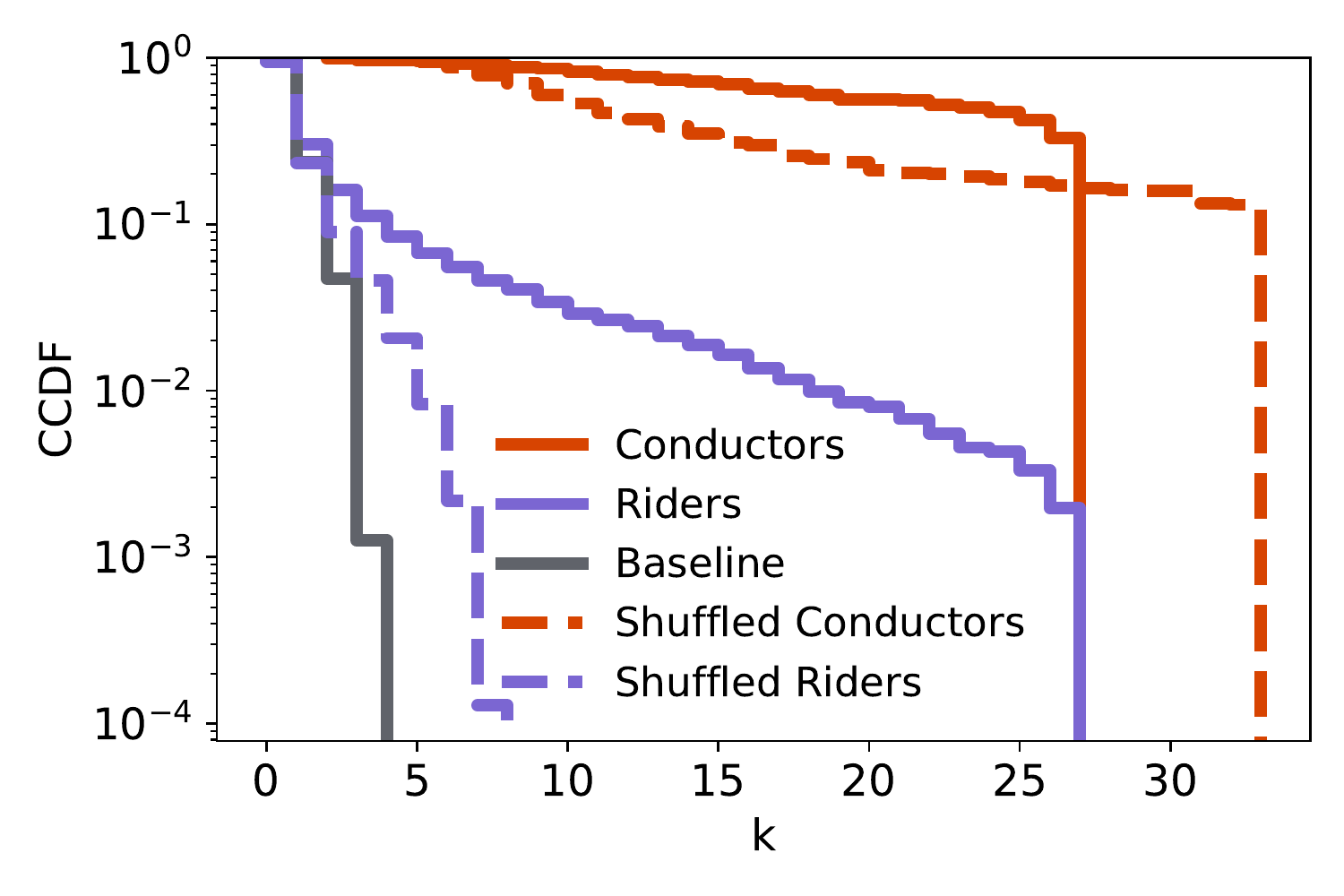}
    \caption{Complementary cumulative distribution of  core number $k$ for accounts in the \dataset{train-net} network and \dataset{tagging-net} baseline, along with a shuffled version of the \dataset{train-net} network.
    KS tests show that all distributions are  significantly different from each other ($p<0.01$).}    \label{fig:kcore}
\end{figure}

We observe that the pro-Trump community in Figure~\ref{fig:mentioning_network} has a densely connected core and many peripheral nodes, while the \dataset{tagging-net} network displays a more homogeneous structure.
To confirm this observation, we perform core decomposition of undirected versions of both networks and calculate the core number of each node, which measures node centrality and influence~\cite{kitsak2010identification}.  Figure~\ref{fig:kcore} plots the distributions of core number $k$.
For the \dataset{tagging-net} baseline network, all nodes have $k$ values smaller than five.
The \dataset{train-net} network, on the contrary, has a deeply hierarchical structure with a very dense core (high $k$). Conductors tend to have higher core values than riders, indicating that they tend to be situated near the core of the network.

Nodes with high degree tend to have high core values. 
To disentangle the roles of degree and $k$ values, we shuffle the edges of the \dataset{train-net} network in Figure~\ref{fig:mentioning_network} while preserving the degrees of the nodes and perform core decomposition again on the shuffled network.
The core number distributions after shuffling are also shown in Figure~\ref{fig:kcore}. 
For the conductors, we observe that high core numbers after shuffling are even larger --- in fact, we can see some high-degree nodes near the periphery of the network in Figure~\ref{fig:mentioning_network}. This suggests that their position near the core of the network is consistent with their high mentioning activity. 
For riders, on the other hand, high core numbers are not explained by degree, suggesting that these accounts gain centrality through the mentions by the conductors.

Let us examine the clustering structure of the networks.
While the \dataset{train-net} and \dataset{tagging-net} networks have similar densities ($3.0 \times 10^{-4}$ vs. $2.3 \times 10^{-4}$), the former has a much higher average clustering coefficient (0.13 vs. 0.04). 
The high number of triangles in the train network suggests that some accounts act as both conductors and riders.

In summary, the above analyses show that compared to the baseline, the mention network of pro-Trump trains is heavily clustered and hierarchically organized around a dense core of highly active conductors.

\subsection{Follow Network}

We are also interested in the follow network induced by the pro-Trump train accounts.
Since querying the ``follow'' relationship between each pair of accounts is not practical due to Twitter's API rate limit, we adopt a sampling strategy.
We split the pro-Trump accounts in \dataset{decahose} into conductor, rider, and baseline groups to examine the follow relations within and across groups.

Studies of political echo chambers have focused on the segregation between conservative and liberal communities on Twitter~\cite{Truthy_icwsm2011politics}. To explore this phenomenon, we consider an additional sample of anti-Trump accounts as a fourth group. We follow an analogous procedure to the baseline accounts in \dataset{decahose}. However, instead of using hashtags frequently found in descriptions of pro-Trump accounts, we query the historical archive using hashtags frequently found in descriptions of anti-Trump accounts encountered during our snowball crawl (see Data Collection Section).

For the four account groups, we sample 5,000 account pairs within each group and use Twitter's friendship API to check whether the accounts in each pair are following each other.
An account pair is considered to have a follow edge if either account follows the other.
The same procedure is also applied to sampled pairs of accounts across groups. 

\begin{table}
\centering
\caption{Percentages of  account pairs with follow edges within and across groups. ``Conductors,'' ``Riders,'' and ``Baseline'' refer to pro-Trump accounts sampled from the \dataset{decahose} dataset, while ``anti-Trump'' accounts are sampled using anti-Trump hashtags (see text).}
\resizebox{\columnwidth}{!}{
    \begin{tabular}{lrrrr} 
        \hline
        & Conductors & Riders & Baseline & Anti-Trump\\
        \hline
        Conductors & 50.3\% & 28.0\% & 17.0\% & 0\% \\
        Riders &  & 8.2\% & 3.8\% & 0\%\\
        Baseline &  &  & 2.8\% & 0\%\\
        Anti-Trump &  &  &  & 2.0\%\\
        \hline
    \end{tabular}
}
\label{tab:following_net}
\end{table}

We report the percentages of account pairs with follow edges within and across groups in Table~\ref{tab:following_net}.
The anti- and pro-Trump  (baseline) groups are clearly segregated, with no cross-connections. What is more interesting is the presence of denser echo chambers within the pro-Trump community, driven by train networks. In particular, riders are more likely to connect to each other than to baseline pro-Trump accounts. And they are even more likely to connect to conductors. Finally, the conductors are more densely connected to baseline and rider accounts, and have highest likelihood to follow other conductors. 
These results confirm the role of follow trains in boosting the clustered structure of pro-Trump echo chambers. 

\section{Profile and Behavioral Characterization}

\subsection{Follow Manipulation}

\begin{figure}[t]
    \centering
    \includegraphics[width=\columnwidth]{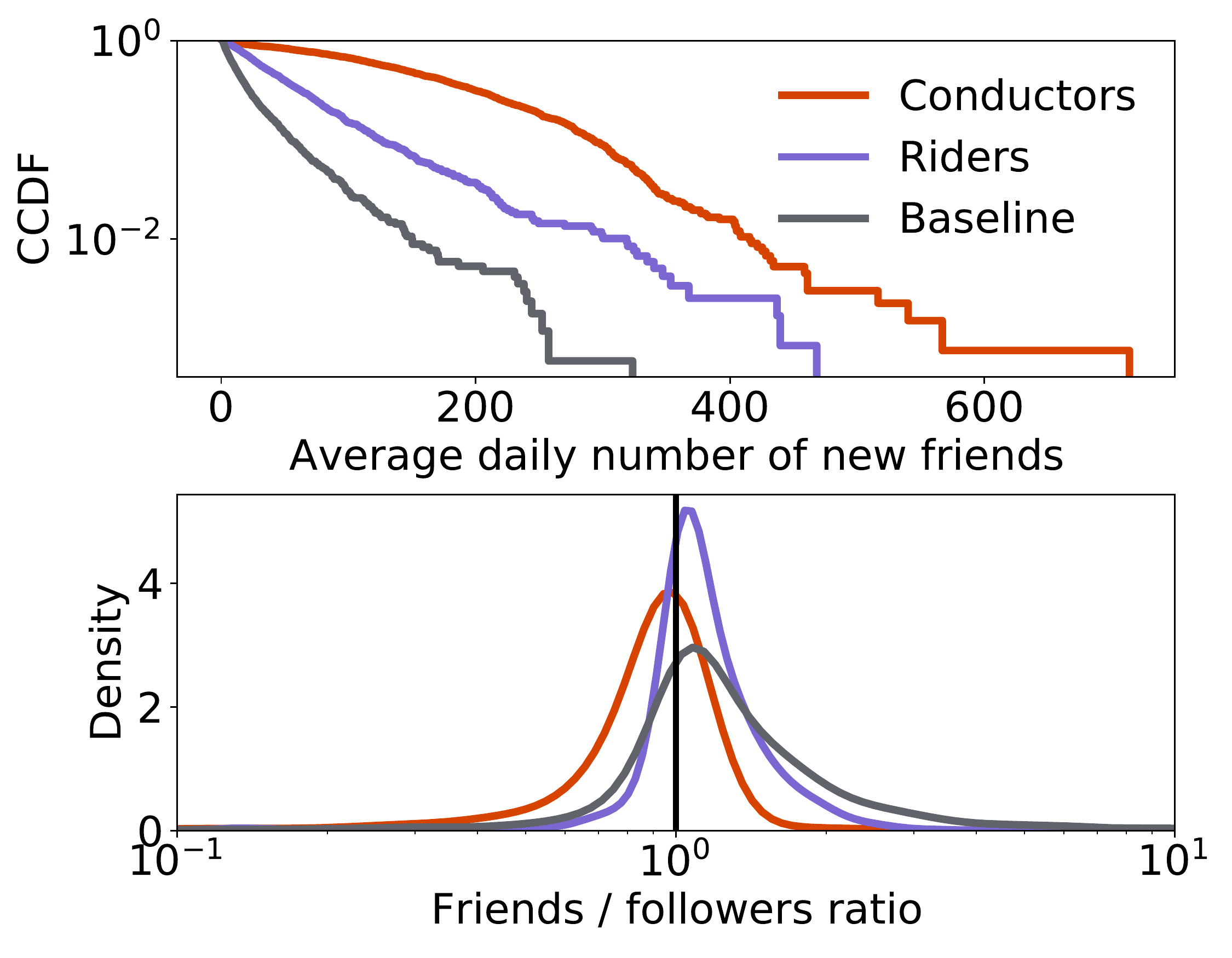}
    \caption{Top: Complementary cumulative distributions of average daily number of new friends for accounts in the \dataset{decahose} dataset. 
    We excluded a few cases with a daily growth above the maximum allowed by Twitter, which may be due to API data errors.
    Bottom: Kernel Density Estimations (KDE) of the distributions of the friend/follower ratios. The vertical black line indicates the ratio of one.
    All distributions are significantly different (KS tests, $p<0.01$).}
    \label{fig:ff_statuses}
\end{figure}

Let us examine the follow behavior of train accounts.
In the \dataset{decahose} dataset, each tweet contains profile information that reflects the status of the author account at the time the tweet was posted.
A series of tweets by the same account provide snapshots capturing the temporal evolution of the account's profile.
By examining the difference in friend counts between tweets in consecutive days, we can estimate the account's daily growth in the number of friends.
We plot the distributions of the average daily number of new friends for different groups of accounts in Figure~\ref{fig:ff_statuses}~(top). Train accounts establish social ties much more aggressively than the baseline.

It important to note that when an account has more then 5,000 friends, Twitter imposes constraints on their friend/follower ratio to prevent manipulation of follow relationships. Although the exact ratio threshold is not published, it is generally understood that an account must have a friend/follower ratio below a critical value close to one. In other words, one can follow additional accounts only after having a similar number of followers. The mutual following patterns promoted by partisan trains are design to circumvent this constraint: riders accumulate followers in order to aggressively follow new accounts. In fact, Figure~\ref{fig:ff_statuses}~(bottom) shows that rider accounts have a friend/follower ratio that is narrowly distributed around one. 

\begin{figure}[t]
    \centering
    \includegraphics[width=0.9\columnwidth]{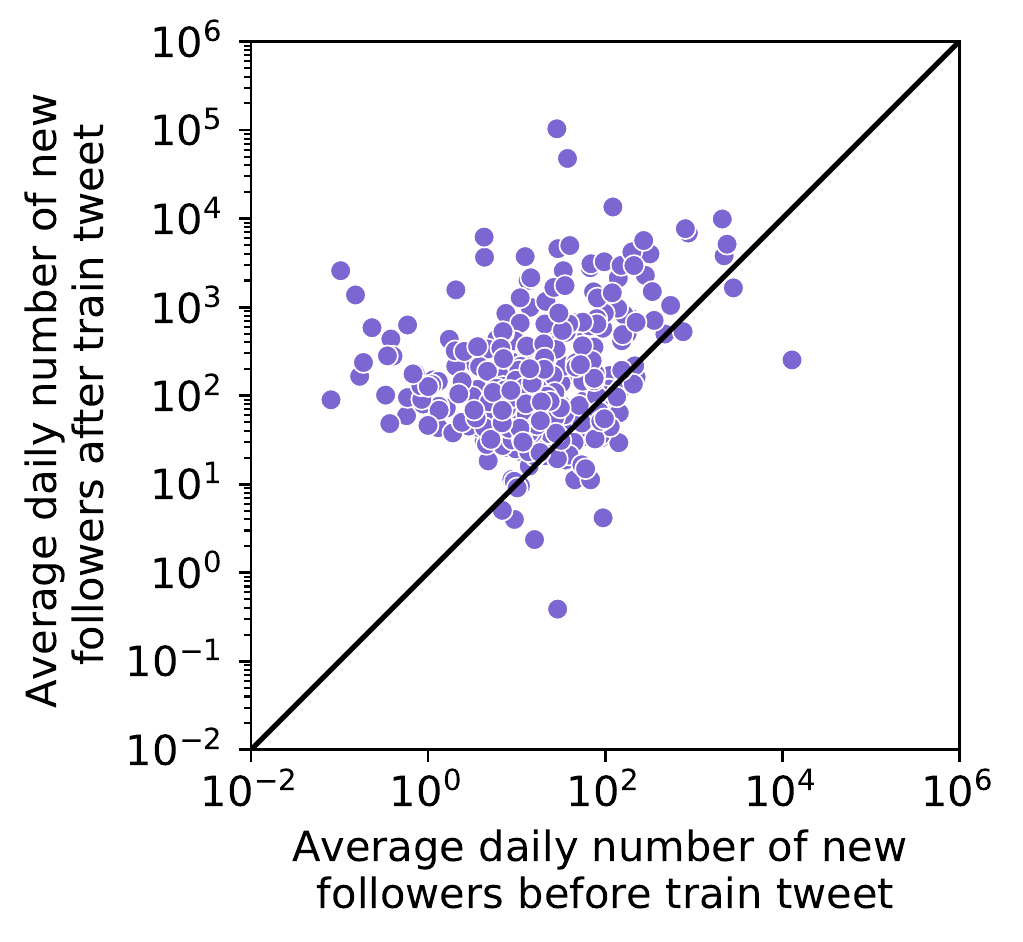}
    \caption{Daily number of new followers of rider accounts. The scatter plot compares the growth before (x-axis) and after (y-axis) the train tweet mentioning each rider. The numbers are obtained from rider tweets before and after a train tweet (see text). The diagonal shows the expected daily growth.
    }
    \label{fig:followers_gained}
\end{figure}

Is follow manipulation by partisan trains effective? Let us quantify the amplification in the number of new followers gained by rider accounts through train tweets. We focus on riders that (i) were mentioned by many train tweets in \dataset{decahose}, (ii) are still active, and (iii) tweeted within 24 hours before and 48 hours after being mentioned. 
For each of these riders, we consider only one of the train tweets mentioning it. Let $t_{\text{train}}$ be the timestamp of the train tweet.
We identify two tweets by the rider. The first occurs at time $t_{\text{before}}$ most immediately preceding the train tweet, i.e., minimizing $t_{\text{train}} - t_{\text{before}}$ s.t. $t_{\text{train}} - 24h < t_{\text{before}} < t_{\text{train}}$. Let $f_{\text{before}}$ be the follower count extracted from this tweet. The second rider's tweet occurs at time $t_{\text{after}}$ after the train tweet and is selected by maximizing the follower count, i.e., $t_{\text{after}} = \arg\max_t(f_t \,\text{ s.t. }\, t_{\text{train}} < t < t_{\text{train}}+48h)$. The difference between the follower counts in these two tweets, divided by the time elapsed since the train tweet, is used to estimate the daily number of new followers gained by the rider after its mention in the train tweet: 
\[\Delta_{\text{before}} = \frac{f_{\text{after}} - f_{\text{before}}}{t_{\text{after}} - t_{\text{train}}}.\]
This number is compared with the estimated daily number of new followers gained by the rider before the train tweet: \[\Delta_{\text{before}} = \frac{f_{\text{before}}}{t_{\text{before}} - t_0},\]
where $t_0$ is the rider's creation timestamp.

Figure~\ref{fig:followers_gained} plots $\Delta_{\text{after}}$ versus $\Delta_{\text{before}}$ for 394 riders meeting the above criteria. Accounts situated on the diagonal line have no change in follower growth. Most of the riders are above the diagonal line (significant per Mann-Whitney U test, $p<0.01$), suggesting that they profit from an increased growth in followers after the train tweet mention. The ratio $\Delta_{\text{after}} / \Delta_{\text{before}}$ indicates a median amplification of the follower gain over 600\%.

\subsection{Accounts Suspension}

Since the 2018 U.S. midterm election season, Twitter implemented more aggressive enforcement actions against policy violations and suspended accounts at a higher rate.\footnote{\url{transparency.twitter.com/en/reports/rules-enforcement.html}}
As the accounts involved in follow trains, especially the conductors, are violating platform policies, we are interested in their suspension rates.
We checked the profile status of all accounts in \dataset{train-net} and \dataset{tagging-net} through the Twitter API on January 6, 2021. (Note that the \dataset{decahose} dataset is not suitable for this analysis because suspended and deleted accounts are removed from the historical archive.) 

Since the original data collection in February 2020, 1,453 train accounts (15\%) had been suspended and 582 (6\%) deleted. 
Breaking these numbers by account type, 22.6\% of the conductors and 14.8\% of the riders were suspended; 7.7\% of the conductors and 6\% of the riders were deleted.
By comparison, 1,334 (24\%) of the accounts in the \dataset{tagging-net} baseline were suspended and 529 (9.5\%) deleted since data collection in March 2020. 
These numbers suggest that train accounts have similar chances to be suspended as the non-political baseline.

\section{Abusive Behaviors}

In this section, we turn our focus to certain automated and abnormal behaviors of train accounts that may flag abuse.

\subsection{Automation}

As discussed in the Background section, various inauthentic actors might be involved in partisan follow trains; we focus on social bots here. 
To estimate the prevalence of automation, we adopt BotometerLite,\footnote{\url{botometer.osome.iu.edu/botometerlite}} a scalable off-the-shelf bot detection tool~\cite{yang2020scalable}. 
For each account, BotometerLite generates a bot score between 0 and 1 with higher values indicating more bot-like behaviors.

\begin{figure}
    \centering
    \includegraphics[width=\columnwidth]{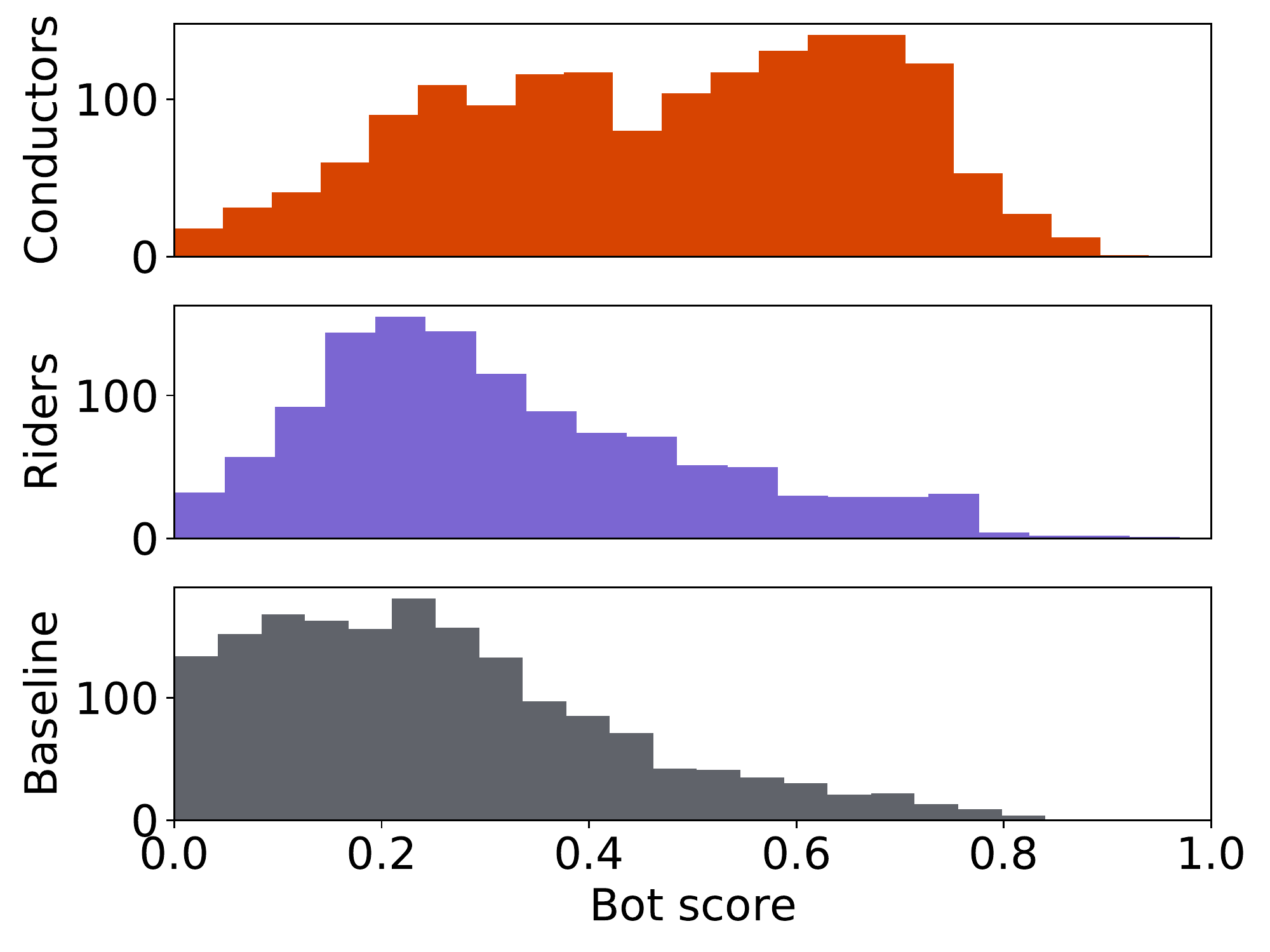}
    \caption{
    Bot score distributions of 
    (top) conductor, (middle) rider, and (bottom) baseline accounts from the \dataset{decahose} dataset. 
    The percentage of bot-like accounts (bot score above 0.5) is 39.4\% for conductors, 12.0\% for riders, and 5.3\% for the baseline.
    The distributions are significantly different (KS tests, $p<0.01$).
    }
    \label{fig:botscore_dist}
\end{figure}

The bot score distributions in Figure~\ref{fig:botscore_dist} show that most baseline accounts in the \dataset{decahose} dataset are human-like.
However, we do observe a larger number of bot-like behaviors among train accounts and especially conductors, suggesting that follow trains may be sustained in part by social bots. 

\subsection{Abnormal Deletion Behaviors}

An examination of the user timelines of some of the train accounts shows that they publish a significant volume of tweets. 
In addition to this, we noticed that some of the train accounts routinely delete their tweets in bulk.
Users have the freedom to delete their posts and may have legitimate reasons to do so~\cite{almuhimedi2013tweets}. 
However, the deletion feature can also be abused. 
For example, trolls and malicious bots may delete their tweets to conceal their activities and intentions and evade detection~\cite{zannettou2019disinformation,yang2019arming,elmas2019lateral}. 

Let us use the \dataset{decahose} dataset to quantify the tweet publishing and deletion events.
Each tweet contains an associated user object with information that reflects the status of the account when the tweet was posted.
Although the data only contains samples of the tweets from each account, these can still provide multiple snapshots of an account at different times.
A decrease or increase in the tweet count between snapshots of an account in consecutive days indicates tweet deletion or new published tweets, respectively. Note that the estimates obtained through this method provide a \emph{lower bound} on the number of new tweets and deletions; the true numbers could be much larger, as a user could post and delete many tweets between consecutive snapshots. 

\begin{figure}
    \centering
    \includegraphics[width=\columnwidth]{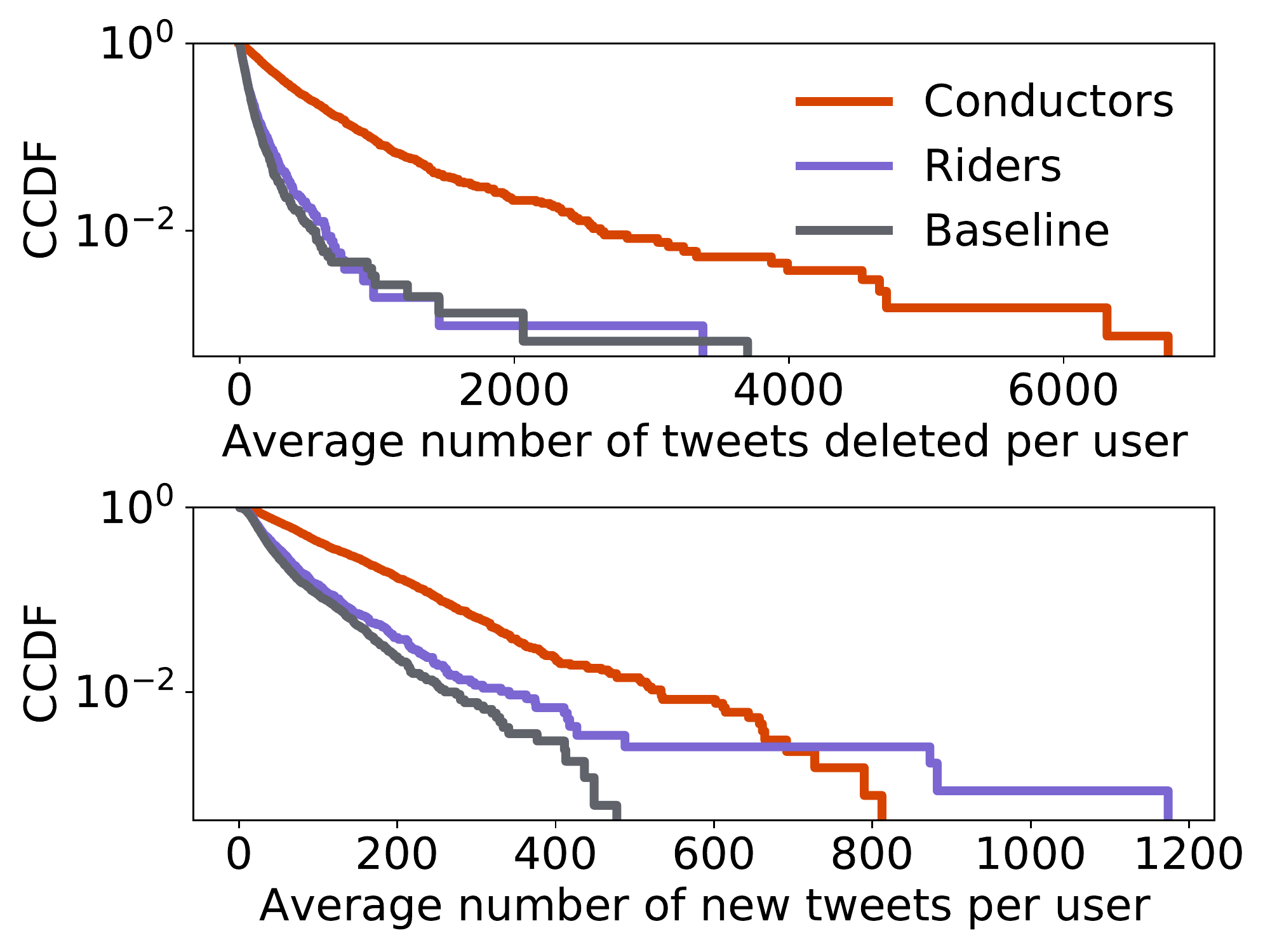}
    \caption{Complementary cumulative distribution of (top) average estimated daily tweet deletions per account and (bottom) average estimated daily new tweets per account in the \dataset{decahose} dataset.
    The distributions are significantly different (KS tests, $p<0.01$), except between riders and baseline on tweet deletions.}
    \label{fig:tweet_deletion}
\end{figure}

Figure~\ref{fig:tweet_deletion} shows the distributions of the estimated numbers of daily tweets deleted and published by accounts in the \dataset{decahose} dataset. 
Conductor accounts perform tweet deletion at a staggering frequency: on average, they delete at least 420 tweets per day.
In contrast, \citet{almuhimedi2013tweets} show that typical Twitter users delete 1--1.6 tweets per day on average. 
The tweet deletion rates of conductor accounts are extremely high even in comparison to rider and baseline accounts (83 and 73, respectively, deleted tweets per day on average). And the tails of the distributions highlight accounts with thousands of deleted tweets per day --- far exceeding the maximum number of posts per day allowed by the platform. 
Although Twitter terms forbid inspection of the deleted content, such abnormal behaviors are strongly suggestive of abuse. 
Train accounts also produce higher volumes of tweets compared to the baseline.

\section{Spreading Toxic Information}

In this section, we analyze the spread of low-credibility news and conspiracy theories by partisan train accounts.

\subsection{Low-credibility News}

We identify low-credibility news based on sources. This approach is widely adopted in the literature~\cite{shao2018spread,grinberg2019fake,pennycook2019fighting,bovet2019influence,yang2020prevalence}  because labeling at the level of individual articles is not feasible~\cite{Lazer-fake-news-2018}. We use the \textit{Iffy+} list of low-credibility sources.\footnote{\url{iffy.news/iffy-plus/}} 
\textit{Iffy+} merges lists of sites that regularly publish mis/disinformation, as identified by major fact-checking and journalism organizations such as Media Bias/Fact Check, FactCheck.org, PolitiFact, BuzzFeed News, and Wikipedia.

\begin{figure}
    \centering
    \includegraphics[width=\columnwidth]{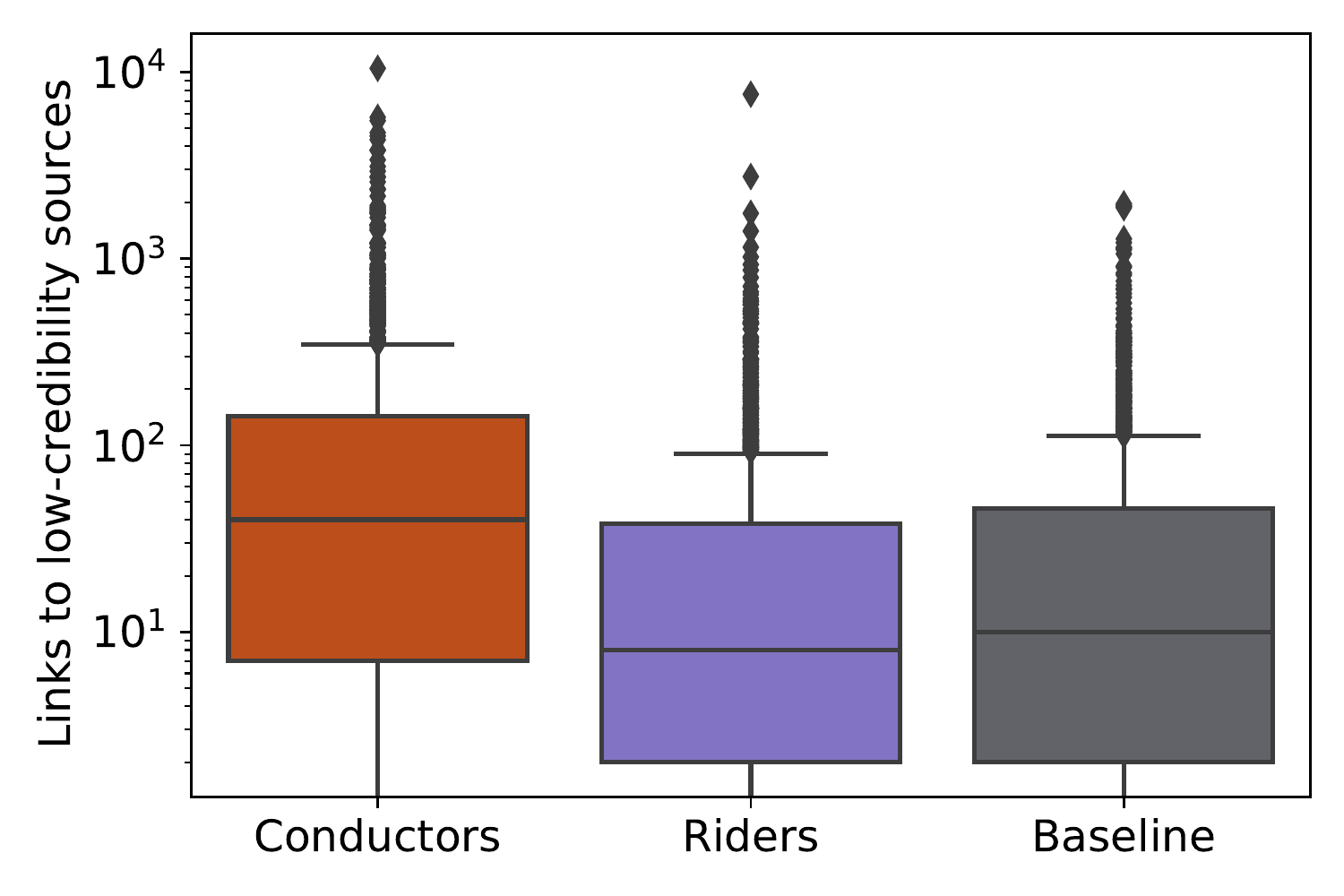}
    \caption{Boxplots showing the distributions of the numbers of links to low-credibility sources shared by accounts in \dataset{decahose}. Conductors share significantly more low-quality links than rider or baseline accounts (U tests, $p<0.01$), whereas the rider and baseline distributions are not significantly different from each other.
    }
    \label{fig:lowcred}
\end{figure}

We find links to low-credibility news outlets embedded in the \dataset{decahose} tweets. 
For each account, we count the \textit{total} number of links to low-credibility sources per user.
Figure~\ref{fig:lowcred} shows that conductors tend to share more of these links compared to rider and baseline accounts. 
Normalizing by account age yields similar results for \textit{daily} numbers of low-quality link shares. 
Part of this difference could be attributed to conductors being more active accounts, which could skew the sample from the historical archive to include more of their tweets.

\subsection{Conspiracy Theories}

\begin{figure}[t]
    \centering
    \includegraphics[width=\columnwidth]{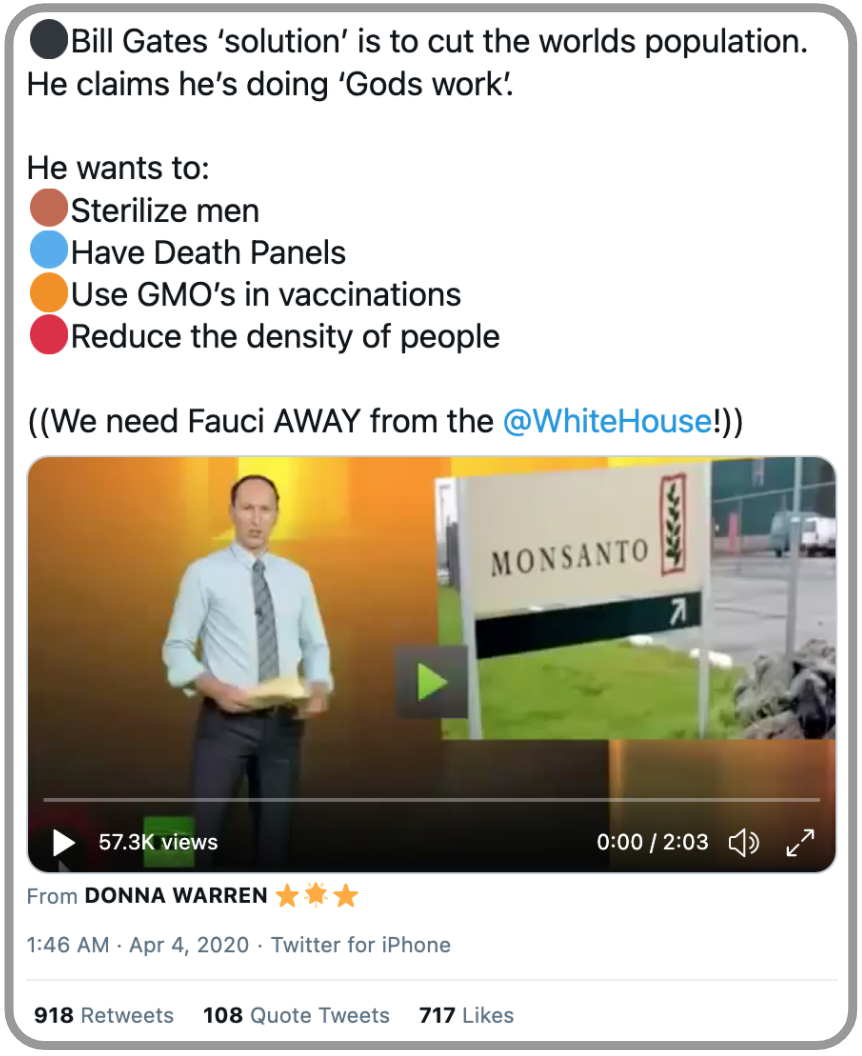}
    \caption{
    Screenshot of a tweet containing conspiratorial claims about COVID-19 vaccines, published by a partisan train account.
    }
    \label{fig:conspiracies}
\end{figure}

Some conspiracy theories have gained significant attention in the context of the 2020 U.S. presidential election and the COVID-19 pandemic.
The content generated and amplified by train accounts includes COVID-19 conspiracy theories, as illustrated by the example in Figure~\ref{fig:conspiracies}.

A particularly notorious and dangerous conspiracy theory is QAnon, which was once only accepted by fringe groups.
As we write this paper, multiple media have reported how QAnon has become more mainstream, merged with false narratives about the pandemic and the election, led to violence, and started to affect people's lives. 
Popular social media platforms, including Facebook\footnote{\url{about.fb.com/news/2020/08/addressing-movements-and-organizations-tied-to-violence}} and  Twitter,\footnote{\url{twitter.com/TwitterSafety/status/1285726277719199746}} recently banned QAnon accounts, pages, and groups.

The results in Figure~\ref{fig:wordcloud} suggest that some of the partisan train accounts label themselves as QAnon believers.
To quantify the involvement of train accounts with QAnon content, we manually curate a list of QAnon keywords (the full list and details are available in the code and data repository). 

\begin{figure}
    \centering
    \includegraphics[width=\columnwidth]{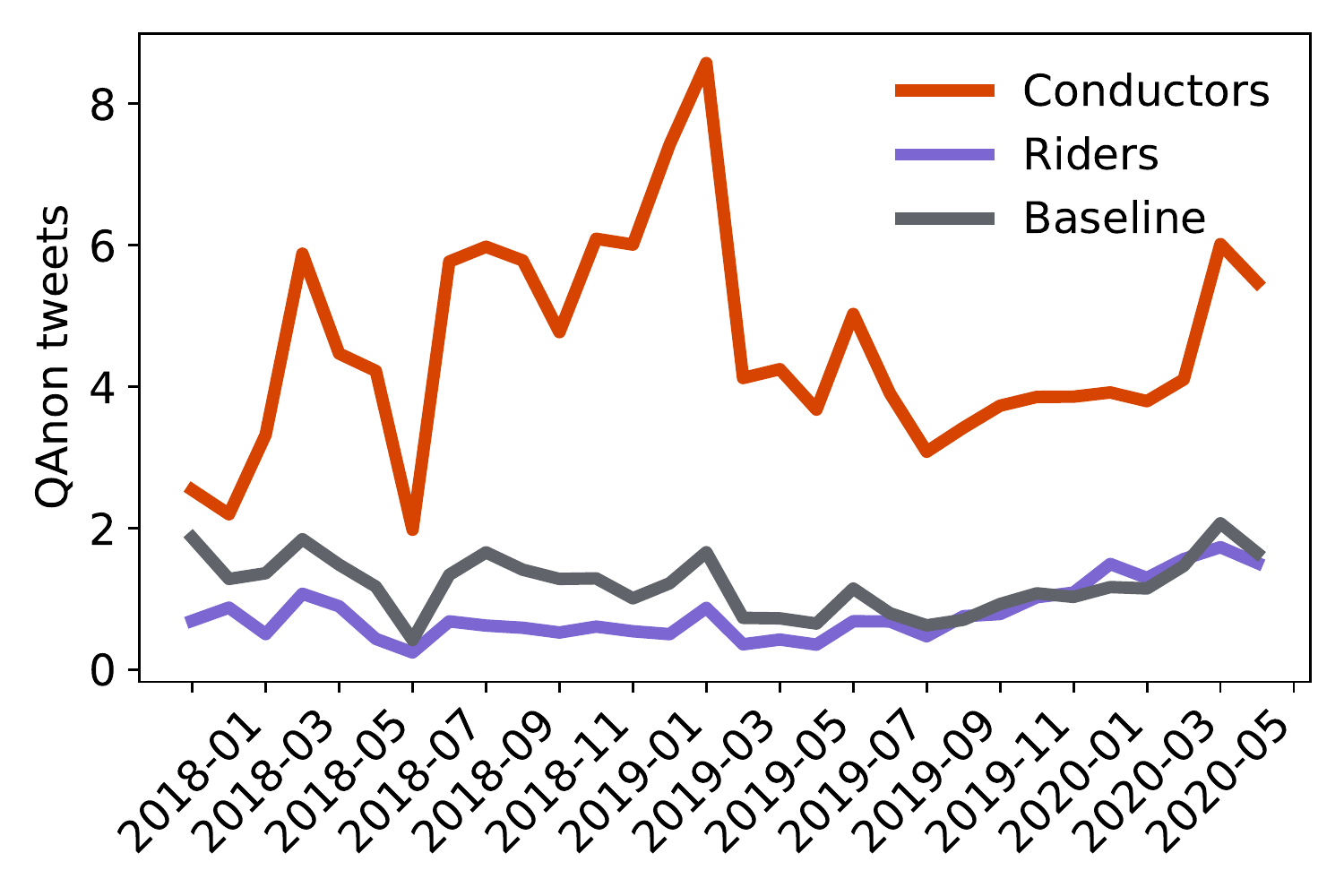}
    \caption{
    Monthly average numbers of \dataset{decahose} tweets per account containing QAnon-related keywords.
    The plot only considers the accounts created until each point in time.}
    \label{fig:qanon_users_twts}
\end{figure}

To check the rate at which QAnon-related content is generated, the keywords are matched against the tweets in the \dataset{decahose} dataset.
We then calculate the monthly average number of QAnon-related tweets for each group and show its time evolution in Figure~\ref{fig:qanon_users_twts}. 
We observe that conductor accounts
produce QAnon content at higher rates than rider and baseline accounts.

\section{Discussion}

This paper provides an in-depth analysis of accounts involved in U.S. partisan follow trains on Twitter. Learning the characteristics of such accounts can help platforms, researchers, policymakers, and the general public recognize follow train abuse and take appropriate measures to curb its harm.

Train accounts manufacture a dense, clustered, and hierarchical echo chamber organized around a small core of conductor accounts. The influence of accounts in this echo chamber is boosted by manipulating follower relationships in ways that circumvent platform rules. 
Train accounts are more likely to display inauthentic and abusive behaviors, such as high-volume posting and deletions, compared to ``ordinary'' partisan accounts with similar descriptions. 
They are also responsible for spreading low-credibility news as well as conspiracy theories. 

Such abusive behaviors negatively affect the online experience of ordinary social media users who are exposed to false and inflammatory information. The real-world consequences are clearly demonstrated by the January 2021 attack on the U.S. Capitol, fueled by a systematic spread of election disinformation, such as QAnon conspiracies amplified by train accounts.\footnote{\url{https://www.poynter.org/fact-checking/2021/a-man-wearing-a-buffalo-cap-proves-how-far-mis-disinformation-can-go-and-how-dangerous-it-can-be/}}

The partisan train phenomenon poses new challenges to social media platforms.
Moderation is needed to mitigate the undesirable outcomes of partisan trains.
Although Twitter has stepped up their efforts to maintain a healthy online discussion around critical issues like elections and public health, our findings suggest that aggressive actions have not yet been taken to curb this particular type of abusive behavior.
For example, Twitter mentions in their Following FAQ\footnote{\url{help.twitter.com/en/using-twitter/following-faqs}} that inauthentic follows by third-party apps can result in account suspension. 
However, the same behavior  by conductor accounts --- whether perpetrated by apps or manually --- is not leading to the prompt suspension of rider accounts.
We believe that moderation policies could be broadened to target abusive follow train strategies. 

Although the present study focuses on the pro-Trump follow train network on Twitter, our data also reveals the existence of an anti-Trump train network.
Future studies should compare the two networks.
Partisan trains also exist on other platforms like Facebook and Instagram, in other countries, and different languages.
Our framework could be applied to extend the present analysis to different platforms and contexts, provided that data from such platforms is available.

\fontsize{9.0pt}{10.0pt}
\selectfont



\bibliography{ref}

\end{document}